\documentclass{aa}
\usepackage{epsfig} 
\usepackage{graphicx}
\usepackage{amssymb}
\usepackage{natbib}
\usepackage{txfonts}
\def\F{F8 \rm{IV}}
\def\XMM{{\em XMM-Newton}}
\def\heao{{\em HEAO 1}}
\def\rosa{{\em ROSAT}}
\def\chan{{\em Chandra}}
\def\lick{{\em Lick Observatory telescope}}
\def\mcg{{\em McGraw Hill telescope}}
\def\kpno{{\em KPNO telescope}}
\def\einstein{{\em Einstein}}
\def\louis{{\em Louisiana State Telescope}}
\def\2129{4U\,2129+47}
\def\x18{X1822-371}

\begin{document}

\title{X-Ray Eclipse Time Delays in \2129}

\author{E. Bozzo\inst{1,2}\fnmsep\thanks{\email{bozzo@oa-roma.inaf.it}} 
\and M. Falanga\inst{3}
\and A. Papitto\inst {1,2}
\and L. Stella\inst{1}
\and R. Perna\inst{4}
\and D. Lazzati\inst{4}
\and G. Israel\inst{1}
\and S. Campana\inst{5}
\and V. Mangano\inst{6}
\and T. Di Salvo\inst{7}
\and L. Burderi\inst{8}
}

\offprints{E. Bozzo}
\titlerunning{\XMM\ observation of \2129}  
\authorrunning{E. Bozzo, et al.}

\institute{INAF - Osservatorio Astronomico di Roma, Via Frascati 33,
00044 Rome, Italy
\and 
Dipartimento di Fisica - Universit\`a di Roma Tor Vergata, via
della Ricerca Scientifica 1, 00133 Rome, Italy
\and 
CEA Saclay, DSM/DAPNIA/Service d'Astrophysique (CNRS FRE
2591), F-91191, Gif sur Yvette, France
\and
JILA, University of Colorado, Boulder, CO 80309-0440, USA
\and
INAF – Osservatorio Astronomico di Brera, via Emilio Bianchi 46, I-23807 Merate (LC), Italy
\and
INAF – Istituto di Astrofisica Spaziale e Fisica Cosmica Sezione di Palermo, via Ugo La Malfa 153, I-90146 Palermo, Italy
\and
Dipartimento di Scienze Fisiche ed Astronomiche, Universit\`a di Palermo, via Archirafi 36 - 90123
Palermo, Italy
\and
Universit\`a degli Studi di Cagliari, Dipartimento di Fisica, SP Monserrato-Sestu, KM 0.7, 09042
Monserrato, Italy
             }

\date{Received 9 August 2007;   
       Accepted  10 September 2007}

    \abstract {} {\2129\ was discovered in the early 80's and classified as an
   accretion disk corona source due to its broad and partial 
X-ray eclipses. The 5.24 hr binary orbital period was inferred from the 
X-ray and optical light curve modulation, implying a late K or M spectral type companion 
star. 
The source entered a low state in 1983, during which the optical modulation disappeared 
and an \F\ star was revealed, suggesting that \2129\ might be part of a triple system.  
The nature of \2129\ has since been investigated, but no definitive conclusion 
has been reached.} 
{Here, we present timing and spectral analyses of two \XMM\  
observations of this source, carried out in May and June, 2005.}
{We find evidence for a delay between two mid-eclipse epochs measured $\sim$22 days apart, 
and we show that this delay can be naturally explained as being due to the 
orbital motion of the binary \2129 around the center of mass of a triple system. 
This result thus provides further support in favor of the triple nature of \2129.}
{}  


\maketitle

\section{Introduction}
\label{sec:intro} 
\begin{table*}[ht!]
  \centering
\caption{Mid-eclipse epoch measurements.}
\begin{tabular}{lccc}
\hline
\hline
\noalign{\smallskip} 
Observatory & Mid-eclipse Epoch$^a$ (JD) & Orbital Period$^a$ (s) & References\\
\noalign{\smallskip}
\hline
\noalign{\smallskip} 
\heao\ and \lick\ & $2443760.755(3)$ & $18857(3)$ & \citet{Th79} (Th79)\\
\hline
\noalign{\smallskip}  
\mcg\ and \kpno\ & $2444107.785(3)$ & $18857.5(1)$ & \citet{mc81} (MC81)\\
\hline
\noalign{\smallskip}  
\einstein\,  & $2444403.743(2)$ & $18857.48(7)$ & \citet{mc82} (MC82)\\
\louis,  &  &  & \\
\mcg\ &  &  & \\
\hline
\noalign{\smallskip}
\chan\ & $2451879.5713(2)$ & $18857.631(5)^{b}$ & \citet{now} (N02)\\
\hline
\noalign{\smallskip}
\XMM\ & $2453506.4825(3)$ & $18857.594(7)$ & this work (tw)\\
\hline
\noalign{\smallskip}
\XMM\ & $2453528.3061(4)$ & - & this work (tw)\\
\hline
\noalign{\smallskip}
\multicolumn{4}{l}{$^{a}$ Numbers in parentheses are the uncertainties on the last significant digit (errors at 1$\sigma$ level).}\\ 
\multicolumn{4}{l}{$^{b}$ Average orbital period calculated by using the two \chan\ eclipses.}  
\end{tabular}
\label{tab:oldobs}
\end{table*} 
\2129\ was discovered by \citet{form}
at a flux level variable between $2.4$ and $4.8\times 10^{-10}$ 
erg cm$^{-2}$ s$^{-1}$ (2--10 keV band). 
Observations of \2129\ in the early $80$'s  showed that both its X-ray 
and optical light curves were modulated 
over a 5.24 hr period, with a partial V-shaped minimum maintaining     
approximately the same shape and phase (\citealt{Th79}; \citealt{ulm}; \citealt{mc82}, hereafter MC82). 
A late K or M spectral type companion of $\sim$0.6 M$_{\odot}$ was suggested, assuming it filled 
its Roche lobe, and the discovery of a type I X-ray burst (\citealt{gar87}) 
led to the classification of \2129 as a neutron star (NS) low mass X-ray binary (LMXB) 
system (\citealt{Th79}; \citealt{mc81}; MC82). 
The source distance was estimated to be $\sim$1--2 kpc, corresponding to an X-ray luminosity 
of $\sim$5$\times10^{34}$ erg s$^{-1}$ (\citealt{horne}). 
The optical light curve could be understood in terms of the 
varying viewing geometry of the X-ray heated 
face of the companion, while the V-shaped minimum 
in the X-ray light curve was interpreted as being due to the gradual eclipse of an 
extended accretion disk corona (ADC). 
The shape of the partial X-ray eclipse and the rapidity of 
its ingress and egress have been used to place 
upper limits on the size of this X-ray scattering region 
($\sim$5$\times10^{10}$ cm for the \2129\ high luminosity state, MC82).  
The origin of ADCs is not well understood yet, 
but it is most likely related to systems in which the mass accretion rate 
is sufficiently high that a tenuous scattering corona is formed as a consequence 
of matter evaporation from the accretion disk (\citealt{wh82}).

\2129\ was first revealed in a low state (F$_{0.3-6~ \rm keV}$$\lesssim$10$^{-12}$ erg 
cm$^{-2}$ s$^{-1}$) in September 1983 (\citealt{pietsch86}; \citealt{wen83}).  
Optical observations carried out between 1983 and 1987 showed a flat 
light curve without any evidence for orbital modulation,  
while the spectrum displayed features fully compatible with a late type 
\F\ star (\citealt{kal88}; \citealt{chev}). 
The hypothesis of a foreground or a background star seemed unlikely, 
due to the low probability ($\lesssim$10$^{-3}$) of chance superposition.   
This led to the suggestion that \2129\ is part of a triple system (\citealt{Th88}).            
The revised estimate of the source distance was $\sim$6.3 kpc (\citealt{cowley}; \citealt{deut}). 

Hints of a possible detection of a dynamical interaction 
between the F star and \2129 were discussed by 
\citet{gar89} and \citet{cowley}, after the discovery of a $\sim$40 km s$^{-1}$ shift   
in the mean radial velocity measurement derived from the F star spectrum. 
Shifts of this amplitude are indeed expected if the F star is in 
a month-long orbit around the binary (\citealt{gar89}). 
  
\rosa\  and \chan\ observations, carried out between $1991$ and $2000$,  
led to a characterization of the low luminosity state 
of \2129 (\citealt{gar92}; \citealt{gar94}; \citealt{gar99}).   
The refined \chan\ position turned out  
to be coincident with the F star to within $0''.1$ 
(\citealt{now}, hereafter N02) providing support in favor of the triple nature of \2129. 
However, a firm conclusion could not be reached. 
 
Here we report on \XMM\ observations of \2129, and discuss the likely 
detection of a mid-eclipse epoch variation between two 
$\sim$22 day distant observations. We show that this delay is naturally explained as 
being due to the orbital motion of the binary \2129 with respect to the center of mass 
of a triple system. This delay is thus probably the first ``Doppler'' (or, more accurately,  
``Roemer'') X-ray signature of the triple nature of \2129. 
We outline our data reduction procedure in Sect. \ref{sec:observation},  
and present the results of timing and spectral analysis in Sect. \ref{sec:Results}. 
Our conclusions are summarized in Sect. \ref{sec:discussion}.  

\section{Observations and data}
\label{sec:observation}
\XMM\ (\citealt{j01}) observed \2129\ on May 15
and on June 6, 2005 for a total time span of $\sim$80 ks 
(about four orbital periods). 
The total effective exposure time for each observation was
$\sim$13 ks for the EPIC-PN, EPIC-MOS1, and EPIC-MOS2 cameras. The remaining
observing time was discarded due to ground station anomalies and 
high radiation from solar activity filling up of the  
EPIC-PN scientific buffer. 
Heavy contamination due to solar activity resulted in poor 
orbital phase coverage, especially during the first observation.  
Furthermore, the EPIC-PN and EPIC-MOS cameras were found to be unequally affected by
this contamination, thus forcing a different selection of good time intervals 
for the spectral and timing analyses (see Sect. \ref{sec:Results}).  
The observation data files (ODFs) were processed to produce calibrated
event lists using the standard \XMM\ Science Analysis
System (SAS 7.0). We used the {\sc epchain} and {\sc
emchain} tasks for the EPIC-PN and the two MOS
cameras, respectively. Source light curves and spectra were extracted in the
0.2--10 keV band, using circles of $\sim$14.6$''$ radius centered on the
source. This corresponds to $\sim$70\% encircled energy 
\footnote{As described in chapter $3.2.1.1$ (Issue 2.5) of the \XMM\ 
users' handbook.} for both the 
EPIC-PN and EPIC-MOS cameras. Larger circles could not be used 
due to the proximity of the S3--$\beta$ Digital Sky Survey
stellar object (N02). We extracted the background
light curves and spectra from circles of radii $\sim$116$''$ in the nearest
source-free region to \2129.\    
Background and source circles were all
chosen to lie within the same CCD.  The difference in extraction areas
between source and background was accounted for by using the SAS {\sc
backscale} task for the spectra and the {\sc lcmath} task from
{\sc Heasoft} (version 6.1.1) for the light curves.  The average source
count rate was found to be 0.041$\pm$0.001 count s$^{-1}$ in the EPIC-PN 
and 0.010$\pm$0.001 count s$^{-1}$  
in the two EPIC-MOS cameras (errors are 1$\sigma$). 
Given the short effective exposure time, the low count rate of the EPIC-MOS1 
and EPIC-MOS2 cameras did not contribute significantly to the spectral analysis; 
therefore in Sect. \ref{sec:spectrum} 
we discuss only the spectrum from the EPIC-PN camera.

\section{Results}
\label{sec:Results}

\subsection{Orbital ephemerides and eclipse parameters}
\label{sec:ephemeris}
In the \XMM\ EPIC-PN light curves, two eclipses were clearly detected,
one in each pointing. These eclipses were also 
unambiguously found in the EPIC-MOS1 and EPIC-MOS2 data. 
In order to avoid any non-synchronicity problems between the three EPIC cameras, 
we followed method III of \citet{bar}.  
In accordance with this method, all the source and 
background time series of the same observation were extracted within the time 
interval around the eclipse that was found not to be  
interrupted by the presence of a previously removed solar flare (Sect. \ref{sec:observation}).  
In all cases, the time interval selection was carried out by filtering each light 
curve with the {\sc evselect} (version 3.59) keywords ``timemin'' and ``timemax''. 
This additional reduction of the effective exposure time  
(in addition to the one described in Sect. \ref{sec:observation})  
was especially restrictive for the first observation, 
for which a total exposure time of only $\sim$4 ks 
around the eclipse could be used.  
The times of all light curves were corrected to the barycentre of the Solar System 
with the SAS {\sc barycen} task (version 1.17.3), and summed up in each observation 
with the {\sc lcmath} task, in order to maximize 
statistics and thus improve the accuracy with which the eclipse parameters 
could be determined.
To estimate the mid-eclipse times, the light curves 
were rebinned in 300 s bins\footnote{A check was carried 
out \textit{a posteriori} to verify that our results are virtually independent of 
the light binning time.}. 
\begin{figure}[t!]
\centering
 \includegraphics[width=9.0 cm]{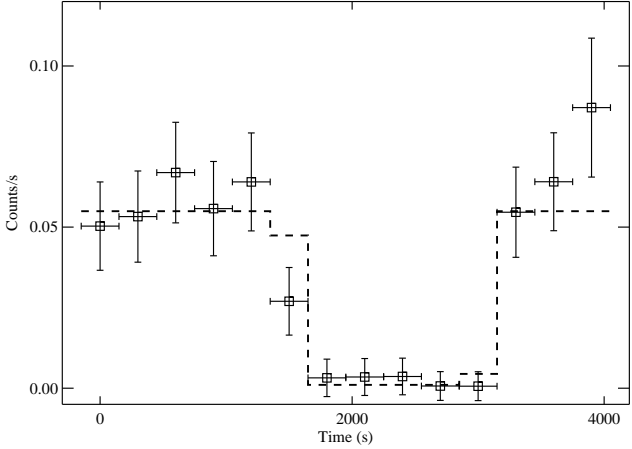}
\caption{
Fit of the mid-eclipse epoch during the first observation. The 0.2--10 keV 
light curve (bin time 300 s) is shown together with the best fit 
model (dashed line).} 
\label{fig:eclipse2}
\end{figure}  

These light curves were then fit with a rectangular eclipse model, in which
the mean count-rate outside ($F_{\rm max}$) and inside
($F_{\rm min}$) eclipse, and the mid-eclipse epoch ($T_0$), were treated as
free parameters. In these fits we fixed the duration of the eclipse 
at the value 1523 s, as measured by N02. 
Being dictated by the secondary star, the duration of the eclipse limb is 
unlikely to have changed since the time of the \chan\ observation 
(see Sect. \ref{sec:spectrum} for details). 
$\chi^2$ minimization was performed with an IDL routine 
written by the authors. 
The model rectangular eclipse was integrated over each time bin
before the $\chi^2$ was computed, in order to take data binning 
into account\footnote{Standard fitting routines that compute
the fitting function punctually in the center of the bin are not
adequate when a function with a large first derivative or features
with a scale smaller than the bin time is considered. In our case the
derivative diverges at the ingress and egress times.}.  
With this method we obtained an accurate determination of the eclipse mid-epoch, 
even though the ingress and egress eclipse times could not be determined 
with the same accuracy.  
The $\chi^2$ hyper-surface was directly sampled in a fine grid of 
values in order to distinguish local minima.
The variance between model and data was then calculated in
each point and for each set of parameters, in order to investigate the
local $\chi^2$ minima in the 4D parameter space.  
The best fits to the eclipse epochs were found to be 
 T$_{\rm 0}$(a)=2453506.4825$\pm$0.0003 JD 
and T$_{\rm 0}$(b)=2453528.3061$\pm$0.0004 JD with $\chi^2/$dof(a)=9/11, 
$\chi^2$/dof(b)=44/42 (errors are at 67\% confidence level unless otherwise specified; 
our epochs are given in UT\footnote{Note that HJD(UT)=HJD(TT)-64.68 s at our epochs. 
We did not consider the correction for the difference between heliocentric Julian date in the 
geocentric (terrestrial) dynamical time system, HJD(TT), and barycentric dynamical time system, BJD(TB).  
The latter is the one used by the SAS {\sc barycen} task, but the difference between 
BJD(TB) and HJD(TT) is less than $\sim$3 s at any given time, which is much smaller than the accuracy 
of our measurement here.}).  
The values of the reduced $\chi^2$ in the above fits are close to 1 and
therefore the addition of any other free parameter in the fit would not 
be justified from a statistical point of view. We also checked that by allowing the eclipse 
duration to vary within the N02's 1$\sigma$ confidence level, i.e., 1473--1553 s,  
the other parameters of the best fit remain unchanged to within the errors  
(mid-eclipse epochs differed by less than 10 s and the 1$\sigma$ errors remained below 26 s). 
We note however that if the eclipse duration is included as a free parameter without any  
constrains, a less significant eclipse delay of $\delta T_{\rm m}$=151$\pm$55 s
would be obtained. 

\begin{figure}[t!]
\centering
 \includegraphics[width=9.0 cm]{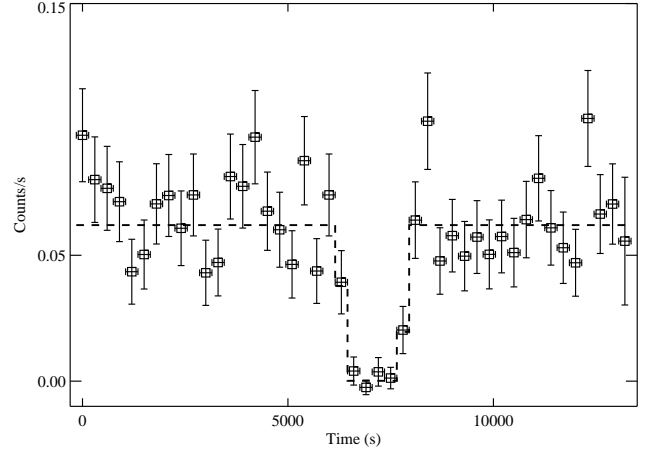}
\caption{
Fit of the mid-eclipse epoch during the second observation. The 0.2--10 keV 
light curve (bin time 300 s) is shown together with the best fit 
model (dashed line).} 
\label{fig:eclipse}
\end{figure}

In the following we adopt values of the two mid-eclipse epochs as derived from the first fit.
We show in Figs. \ref{fig:eclipse2} and \ref{fig:eclipse} the two  eclipses,  
together with the best fit models discussed above.  
We also carried out the fits by using a modified version of the eclipse model by \citet{Mang} 
in order to estimate the eclipse ingress and egress time. 
We allowed these times to take independent values during the fit (N02). 
Only upper limits of $\lesssim$310 s could be derived, 
which are significantly larger than 
those obtained from the \chan\ observation (in 130--260 s range, N02). 

\begin{figure}[t!]
 \centering
  \includegraphics[width=6.4 cm,angle=-90]{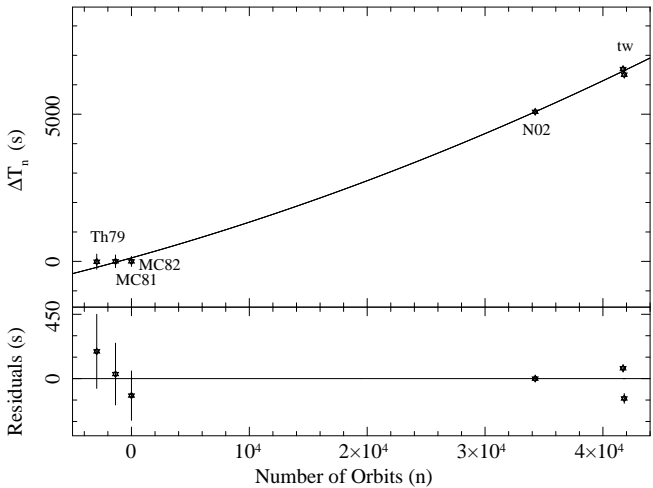}
   \caption{Delays of the mid-eclipse epochs with respect to a constant 
   $P_{\rm orb}$=$P_{\rm ref}$ model. The solid line in the upper panel represents 
   the quadratic best fit to the epochs (see Sect. \ref{sec:ephemeris}). 
  The lower panel shows the residuals from this fit.}     
\label{fig:ephemeris}
\end{figure}

In order to determine a refined orbital solution, 
we considered the above mid-eclipse epochs together with the epochs, 
$T_{\rm n}$, derived from earlier observations (see Table \ref{tab:oldobs}). 
As discussed in Sect. \ref{sec:intro}, the observed phase
alignment between the X-ray and optical light curve minima, allows
the comparison of optical and X-ray  
measurements of the system's ephemerides.
The long time span covered by the eclipse measurements 
(1979--2005 or $n_{\rm max}$$\sim$44800) 
can be used to improve the accuracy of the orbital solution and, 
possibly, measure the orbital period
derivative. 
To this aim we used a standard O--C technique\footnote{Observed minus calculated 
residuals, which are the delay in eclipse time over that expected for a 
constant period system (see, e.g., \citealt{parmar91}; \citealt{pap})   
and references therein.}. 
We considered the ephemeris from MC82  
as reference ($T_{\rm ref}$=2444403.743$\pm$0.002 JD, 
$P_{\rm ref}$=18857.48$\pm$0.07 s), and 
plotted in Fig. \ref{fig:ephemeris} the delays 
$\Delta T_{\rm n}$=$T_{\rm n}$-$T_{\rm n_{\rm pred}}$. Here  
$T_{\rm n_{\rm pred}}$=$T_{\rm ref}$+n$P_{\rm ref}$, 
with n the closest integer to ($T_{\rm n}$-$T_{\rm ref}$)/$P_{\rm ref}$ 
(our two observations correspond to n=41706, 41806). 
In the same figure we have also plotted the best quadratic fit to the O--C  
residuals, corresponding to an orbital period evolution with constant time derivative. 
A linear fit (i.e., a constant orbital period) 
to the same data gave an unacceptable fit ($\chi^{2}$/dof=91/4). 
Table \ref{tab:ephemeris} gives our corrected reference time, ${T_{\rm ref}}$, 
orbital period, ${P_{\rm ref}}$, and the derived orbital period evolution 
${P}_{\rm orb} {\dot{P}}_{\rm orb}^{-1}$=(5.8$\pm$0.7)$\times10^6$ yr.  
This value is a factor of $\sim$4 larger than that in N02, but 
we note that N02's estimate was deduced by adopting the MC82 value of 
$P_{\rm orb}$ and accounting for the entire measured delay as being due 
to an orbital period derivative.  
The observed delay of $\sim$6500 s between the 
ephemeris of MC82 and the one found in the present work 
(see Fig. \ref{fig:ephemeris}), implies an orbital phase shift of $\sim$0.35. 
Small deviations in the eclipse centroid between the active 
and quiescent state of \2129\, may be introduced by changes in the 
shape of the ADC around the compact object. 
However a $\sim$6500 s delay is far too large to be 
explained as a result of such geometrical variations. 
The poor $\chi^{2}$ in Table \ref{tab:ephemeris}  
is due to the large shift of the first \XMM\ point  
with respect to the second one ($\delta T_{\rm m}$=192$\pm$43 s), which could not 
be accounted for by any quadratic fit to the eclipse phase evolution. 
While this delay is much smaller than that discussed above, 
we argue that it is also very unlikely to result from 
geometrical variations within the \2129\ binary system. 
This is because the eclipse profile is consistent with the central source being 
eclipsed by the companion star. Moreover, 
the source X-ray flux and spectrum remained virtually the same across 
the two \XMM\ observations and the \chan\ observation discussed by N02 
(see Sect. \ref{sec:spectrum}).  
In Sect. \ref{sec:discussion} we discuss the possibility that this 
delay is due to light propagation in a triple system.\\ 

We also extracted the EPIC-PN 
light curves of the observations, 
since these have a better orbital phase coverage than 
those obtained by summing all three instruments. This however resulted in 
a lower count rate and S/N. 
\begin{table}[b]
\centering
\caption{Orbital solution obtained with the best quadratic fit 
to the O--C delays $\Delta T_{\rm n}$ (see Fig. \ref{fig:ephemeris}).}
\begin{tabular}{ll}
\hline
\hline
\noalign{\smallskip}
${T_{\rm ref}}$ (JD) & 2444403.7443$\pm$0.0013 \\
${P_{\rm orb}}$ (s) & 18857.594$\pm$0.007\\
${\dot{P}_{\rm orb}}$ (s s$^{-1}$) & (1.03$\pm$0.13)$\times$10$^{-10}$ \\
${P_{\rm orb}\dot{P}_{\rm orb}^{-1}}$ (yr) & (5.8$\pm$0.7)$\times$10$^{6}$ \\
$\chi^{2}$/dof & 25.6/3 \\
\hline
\end{tabular}
\label{tab:ephemeris}
\end{table} 
These light curves and those obtained by using data summed over the three EPIC cameras  
were folded at the best orbital solution using 10 
phase bins. We fitted these two light curves with the function 
F($\phi$)=A+B$\sin$[2$\pi$($\phi$-$\phi_0$)], and looked for a  
sinusoidal modulation similar to that observed by N02. 
No significant modulation was observed (fitting with a constant
value gave a $\chi^2$/dof of 12/13 and 10/17 respectively).     
We derived a 90\% confidence upper limit on the 
amplitude modulation of $\sim$17\%, i.e.,   
a factor of $\sim$2 smaller than the value reported by N02.  
This result was also checked by using light curves extracted only 
in the 0.5--2.0 keV band, where the amplitude modulation might be 
higher (N02). No significant differences were found.     
We discuss these results in Sect. \ref{sec:discussion},
together with the results from the spectral analysis
(Sect. \ref{sec:spectrum}).
\begin{table*}[ht!]
\begin{center}
\caption{\label{table:spec}Best-fit spectral parameters with $N_{\rm H}$ and {\sc bb} model (90\% confidence level error bars).}
\begin{tabular}{lccllcc} 
\hline
\hline 
\noalign{\smallskip}  
&1° Obs.&(May-15-2005)~~~~~~ &&2° Obs.&(June-6-2005)~~~~~~ &\\ 
\noalign{\smallskip}
& $\chi^2$-Stat & C-Stat && $\chi^2$-Stat & C-Stat &\\
\hline
\noalign{\smallskip}
\noalign{\smallskip}
$N_{\rm H}$ (10$^{22}$ cm$^{-2}$)  & $0.14^{+0.11}_{-0.09}$  & $0.24^{+0.03}_{-0.09}$ & & $0.21^{+0.1}_{-0.06}$ & $0.28^{+0.03}_{-0.08}$ &\\  
$kT_{\rm bb}$ (keV) & $0.25^{+0.03}_{-0.03}$ & $0.22^{+0.03}_{-0.02}$ & & $0.20^{+0.02}_{-0.02}$ & $0.18^{+0.02}_{-0.01}$ &\\
$R_{\rm bb}^{\rm a}$ (km) & $1.3^{+1.1}_{-0.4}$ & $1.9^{+0.74}_{-0.53}$ & & $2.42^{+1.8}_{-0.7}$ & $2.87^{+1.7}_{-0.6}$ &\\
$\chi^{2}$/dof & $14.30/14$  & - & & $20.6/23$ & - &\\
C-Stat & - & $496.8$ &  & - & $592.24$ &\\
$F_{\rm 0.2-10 keV}^{\rm a}$ ($10^{-14}$ erg cm$^{-2}$ s$^{-1}$) & 9.3 & 9.3 & & 8.9 & 9.0 &\\
$F_{\rm 0.2-10 keV}^{\rm b}$ ($10^{-13}$ erg cm$^{-2}$ s$^{-1}$) & 1.6 & 2.3 & & 2.2 & 3.3 &\\
$L_{\rm 0.2-10 keV}^{\rm c}$ ($10^{32}$ erg s$^{-1}$) & 7.7 & 10.9 & & 10.5 & 15.8 &\\ 
\noalign{\smallskip} 
\hline  
\noalign{\smallskip} 
\multicolumn{6}{l}{$^{\rm a}$  Absorbed Flux} \\
\multicolumn{6}{l}{$^{\rm b}$  Unabsorbed Flux} \\
\multicolumn{6}{l}{$^{\rm c}$  From the unabsorbed flux and assuming a distance of 6.3 kpc.}\\ 
\end{tabular}  
\end{center}
\end{table*} 
\subsection{Spectral analysis}
\label{sec:spectrum}
\begin{figure}[t!]
\centering
  \includegraphics[width=6 cm,angle=-90]{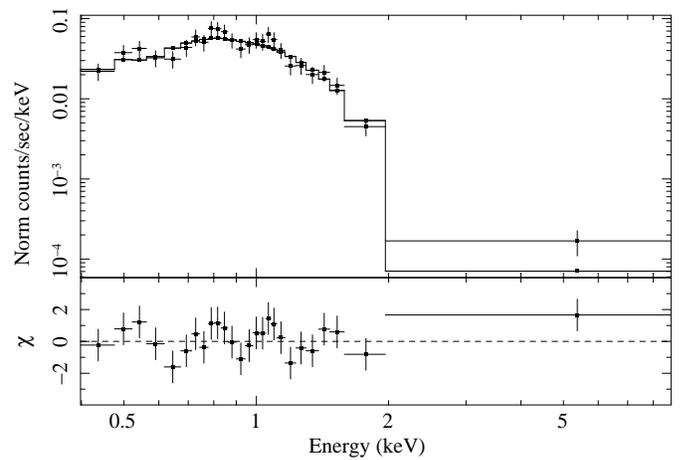}
\caption
{Measured 0.2--10 keV spectrum of \2129\ during the second pointing. 
The best fit model and the contribution of the fit residuals to the $\chi^2$ are also shown.}
\label{fig:spectrum2}
\end{figure}
Spectral analysis was carried out by using XSPEC version
11.3.2t (\citealt{arnaud96}); the data were rebinned in order to have at least 20 photons in 
each energy bin.  Owing to poor statistics, phase-resolved 
spectroscopy could not be carried out. 
Instead, the spectra of the two observations were accumulated 
during the same time intervals selected for the extraction of the EPIC-PN light curves, 
except for the eclipses (see Sect. \ref{sec:ephemeris}). 

The spectrum of the first observation was modeled with only  
an absorbed blackbody (poor statistics did not allow for more complex models).\\ 
The second observation was first modeled with an absorbed power law 
plus a blackbody component, but the F-statistics probability with respect to a simple 
absorbed blackbody model was found to be $\sim$0.2. 
The best fit was then obtained, also in this case, by adopting an absorbed blackbody model.\\ 
A power law component with fixed photon index $\Gamma$=1.1 (N02) 
was added to the fit of the second observation in order to estimate an upper limit. 
For such a power law component, the 90\% confidence upper limit was about 10\% of 
the 0.2--10 keV unabsorbed flux (in agreement with N02, see their model A). 
C-statistics model fitting (\citealt{cash}) was also performed 
on the unbinned spectra: the results were fully compatible with those obtained by 
using $\chi^2$ minimization. 
Figure \ref{fig:spectrum2} shows the spectrum and model of the second pointing 
as an example, while  
the best fit parameters are reported in Table \ref{table:spec}. 
No significant difference was found between the two \XMM\ observations, 
and all parameters were compatible, to within the errors, with those determined by N02 
for the quiescent state of \2129.\  
We also tested our results with the {\sc xspec nsa} model (\citealt{arnaud96}; \citealt{zavlin}). 
Fits were carried out, first by using a fixed distance of 6.3 kpc, 
and then by fixing a neutron star radius of 5 and 10 km (N02). 
Results of these fits were found to agree with those of N02 (our errors on all parameters are a 
factor of $\sim$1.5 larger).    
 
\section{Discussion}
\label{sec:discussion}
We reported on \XMM\ observations of \2129\ in its quiescent state, 
which has lasted, apparently uninterrupted, since 1983 (\citealt{wen83}). 
The discovery of a late F-type star coincident with the position of \2129\ 
(\citealt{Th88}; \citealt{chev}) led to the hypothesis 
that this binary system might be part of a hierarchical triple. 
Our detection of a delay $\delta T_{\rm m}$=192$\pm$43 s across two eclipses 
separated by $\sim$22 days, can be naturally explained as being due to 
the orbital motion of the binary 
with respect to the center of mass of a triple, and thus 
lends support in favor of the triple system hypothesis.  

Using a third star of mass $M_1$, an inner binary with  
$M_2$$\sim$2 M$_{\odot}$, and a non-eccentric orbit, 
the expected delay between two mid-eclipse epochs 
separated by a time interval $\tau$ can be expressed as  
\begin{equation}
\delta T=a_{2}\sin{i} /c \left[\sin{(\phi_0+\delta\phi)}-\sin{\phi_0}\right]. 
\label{deltat}
\end{equation}
Here $a_2$=(G/4$\pi^2$)$^{1/3}$ P$_{\rm tr}^{2/3}$ $M_1$/($M_1$+$M_2$)$^{2/3}$ 
is the radius of the inner binary orbit around the triple system 
center of mass (with a period $P_{\rm tr}$), $\phi_0$ is the 
phase at $T_0$(a) of the binary system along such an orbit, 
c is the speed of light, i is the inclination angle of 
the triple system orbit, $\delta\phi$=2$\pi\tau$ P$_{\rm tr}^{-1}$ 
and, in our case, $\tau$=$T_0$(b)-$T_0$(a)$\simeq$22 d. 

\begin{figure}[t!]
 \centering
  \includegraphics[width=8.0 cm, height=8.5cm]{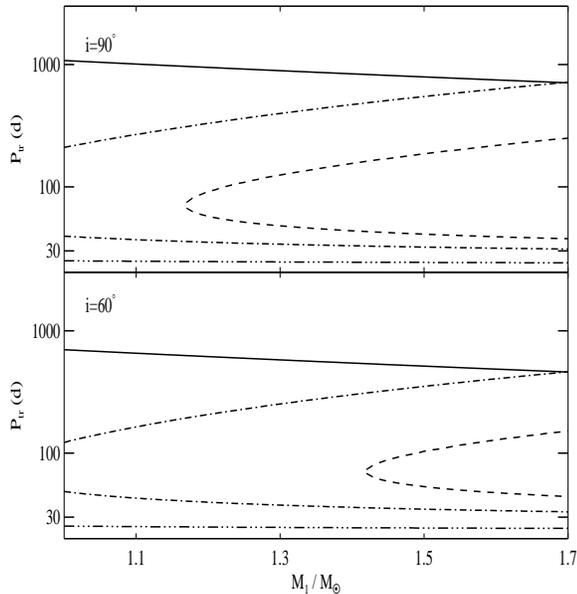}
   \caption{Allowed regions of the triple orbital period, $P_{\rm tr}$, as a 
   function of the third star mass, $M_1$,  
   in the cases in which the measured delay $\delta T_{\rm m}$=192 s is varied within 1$\sigma$ 
   (dashed line), 2$\sigma$ (dot-dashed line) and 3$\sigma$ (dot-dot-dot-dashed line) 
   confidence level. For the 3$\sigma$ range only the lower limit is drawn. 
   Solid lines represent the limits imposed by the measured radial velocity shifts of the F-star  
   (see text for more details). The two panels show the cases  
   i=90$^{\circ}$ and 60$^{\circ}$.}     
\label{fig:limits}
\end{figure}

Figure \ref{fig:limits} shows a plot of the range of allowed values of the triple 
orbital period as a function of the third star mass\footnote{For an F-type star it is expected  
$M_1$$\sim$1--1.6 M$_{\odot}$ (see, e.g., \citealt{erika}).}, which give 
delays compatible with the measured value $\delta T_{\rm m}$ (to  
within the uncertainties at a given confidence level). 
The dashed, dot-dashed and dot-dot-dot-dashed lines represent the   
ranges for 1$\sigma$=43 s, 2$\sigma$=85 s and 3$\sigma$=165 s confidence intervals 
in $\delta T_{\rm m}$, respectively. 
The solid lines give the constraints on $P_{\rm tr}$ imposed by the measured 
$\sim$40 km s$^{-1}$ shift in the mean radial velocity of the F star (\citealt{gar89}), under the 
assumption that this represents the maximum observable radial velocity shift
of its orbit.
The two panels of Fig. \ref{fig:limits} show the cases of i=90$^{\circ}$, 
and  i=60$^{\circ}$ respectively.  
From the upper panel of this figure (i=90$^{\circ}$) it can be seen that, at 1$\sigma$ 
confidence level, we can set a lower limit 
on the F star mass of $\sim$1.2 M$_{\odot}$. Considering a 2$\sigma$ uncertainty 
on $\delta T_{\rm m}$ removes the upper limit on $M_1$ and only the orbital period 
can be constrained (39 d$<$$P_{\rm tr}$$<$200 d, for $M_1$$\simeq$1 M$_{\odot}$). 
We note that a decrease of the inclination angle results in a smaller range of allowed orbital 
periods, while the effect of a non-zero eccentricity (hypothesis not considered 
in our calculation) would have the opposite effect. 

The possibility that \2129\ is in a triple system might also have noticeable  
consequences for our measured orbital period derivative,   
$P_{\rm orb}\dot{P}_{\rm orb}^{-1}$=(5.8$\pm$0.7)$\times$10$^6$ yr (see Sect. \ref{sec:ephemeris}). 
At first sight, this value seems to imply that, as in other known LMXBs 
(e.g., \x18, \citealt{parm}; \citealt{heinz}; \citealt{pnw}) the orbital period of \2129\  
is increasing. This would be contrary to evolutionary 
expectations. In fact, angular momentum losses (such as gravitational wave emission 
and magnetic breaking) in a binary system with an orbital 
period of $\sim$5 hr and a non-degenerate companion would imply a decreasing orbital period (\citealt{verbunt}).  
However, if a modulation in the eclipse arrival time of amplitude $T_{\rm a}$=$a_2$$\sin{i}$/c (as discussed above) 
contributes to the measured epochs of the eclipses, caution should be used with any estimate of 
$P_{\rm orb}\dot{P}_{\rm orb}^{-1}$ (such as that in Sect. \ref{sec:ephemeris}) 
that does not take this modulation into account. 
Given our poor knowledge of the parameters of the triple system, a detailed correction 
for the orbital motion of the inner binary around the center of mass of the triple 
cannot be carried out yet. 
A conservative estimate of $P_{\rm orb}\dot{P}_{\rm orb}^{-1}$ can be 
derived by increasing the uncertainties on the observed $\Delta T_{\rm n}$ 
up to a value $T_{\rm a}$, which represents the unknown 
amplitude of the modulation in the eclipse arrival times.     
The range of values that $T_{\rm a}$ may attain depends mainly on the range of orbital periods 
for which a solution of Eq.{\ref{deltat}}, compatible with the values of $\tau$ and $\delta T_{\rm m}$ 
we measured, exists. 
Unfortunately, the range on $P_{\rm tr}$ that can be inferred from Fig. \ref{fig:limits} is fairly loose. 
Based on our 1$\sigma$ range of $\delta T_{\rm m}$, 
we derive a minimum upper limit of $P_{\rm tr}^{\rm upp}$$\simeq$75 d for the third star orbital period 
(for $M_1$$\simeq$1.2 M$_{\odot}$ and i=90$^{\circ}$), 
which corresponds to $T_{\rm a}$$\simeq$100 s. We checked that increasing  
the uncertainty on the measured values of the eclipse epochs by this value 
would remove entirely the need to introduce  
an orbital period derivative.  
By analogy we speculate that the presence of a third body in a wide orbit 
around the inner binary might explain the positive orbital period 
derivatives observed in some LMXBs, which is at odds with the expectations of the 
standard evolutionary scenario.\\ 

Due to the low count-rate, and short effective exposure time, our 
\XMM\ observations did not detect any spectral component 
above 2 keV. An absorbed ($N_{\rm H}$$\sim$0.2$\times$10$^{22}$ cm$^{-2}$) 
blackbody component with kT$_{\rm bb}$$\sim$0.2 keV and $R_{\rm bb}$$\sim$2 km 
provided an adequate modeling of the spectrum.  

For quiescent NSs in LMXBs, like \2129,\ this soft X-ray emission 
can be produced in several alternative ways. 
One possibility is that this emission is powered by thermal energy released 
as the NS cools in between accretion phases (\citealt{brown}; \citealt{colpi}; \citealt{win}).   
Alternative models invoke a NS in the propeller regime, 
shock emission due to the interaction between the pulsar wind and 
matter in the vicinity of the companion star, and residual accretion 
onto the compact star (see, e.g., \citealt{stella94}; \citealt{campana98}). 
While the properties of the propeller regime are still rather uncertain, 
a power law emission of photon index 1--2 should be expected at 
least in the last two interpretations (\citealt{campana98}; N02). 
Discriminating between these models is not possible based on our results, 
given the lack of information at energies above 2 keV. 

A soft spectral component with similar properties, plus a power law component 
with photon index $\sim$1.1, was observed in the quiescent state of \2129\ in December 2000 (N02). 
However, also this observation was hampered by a low number of counts above 2 keV, 
thus preventing an accurate characterization of the power law emission. 
Some indication was found that the power law component was consistent with 
being of constant amplitude and slope, while the blackbody component was  
sinusoidally modulated over the orbital period, in a manner consistent with 
neutral column density variation (a factor of $\sim$2). 
This modulation was ascribed to the presence of a vertically extended 
disk atmosphere, thicker at the outer rim,    
close to the region where the accretion stream from the secondary star 
impacts. 

As discussed in Sect. \ref{sec:spectrum}, our folded light curve showed no clear 
indication of a sinusoidal modulation.    
The upper limit we derived on the amplitude of this modulation 
is significantly lower than that discussed by N02 (a factor of $\sim$2). 
This suggests that there has been some change in the geometry of the outer disk region 
(in particular the region where the stream from the secondary impacts) across 
the \chan\ and \XMM\ observations. 

A series of monthly spaced \XMM\ or \chan\ observations would afford a much 
more accurate characterization of the quiescent emission of \2129\, and 
measure with good accuracy the modulation in the X-ray eclipse times due to the third body.  
This will yield much needed information on the triple system parameters and 
single out unambiguously any orbital period 
evolution. 

\begin{acknowledgements} 
EB thanks University of Colorado at Boulder and JILA  
for the hospitality during part of this work, 
E. Piconcelli, E. Costantini and N. Rea for useful 
discussions. MF acknowledges the French Space
Agency  (CNES) for financial support.  
This work was partially supported through ASI 
and MIUR grants.
\end{acknowledgements}


\begin{thebibliography}{}

\bibitem[\protect\citeauthoryear{Arnaud}{1996}]{arnaud96} Arnaud, K. A. 1996, ASPC, 101, 17
\bibitem[\protect\citeauthoryear{Barnard et al.}{2006}]{bar} Barnard, R., Trudolyubov, S., Haswell, C. A., et al. 2007, A\&A, 469, 875 
\bibitem[\protect\citeauthoryear{Bohm-Vitense}{1992}]{erika} Bohm-Vitense E., in Introduction to stellar astrophysics, v.3, Cambridge 
University press 1992, p. 16
\bibitem[\protect\citeauthoryear{Brown et al.}{1998}]{brown} Brown, E. F., Bildsten, L., Rutledge, R. E. 1998, ApJ, 504, L95
\bibitem[\protect\citeauthoryear{Campana et al.}{1998}]{campana98} Campana, S., Colpi, M., Mereghetti, S., Stella, L., Tavani, M. 
1998, A\&ARv, 8, 279
\bibitem[\protect\citeauthoryear{Cash}{1979}]{cash} Cash W. 1979, ApJ, 228, 939
\bibitem[\protect\citeauthoryear{Chevalier et al.}{1989}]{chev} Chevalier, C., Ilovaisky, S. A., Motch, C., Pakull, M., Mouchet, M. 
1989 A\&A, 217, 108
\bibitem[\protect\citeauthoryear{Colpi et al.}{2001}]{colpi} Colpi, M., Geppert, U., Page, D., Possenti, A. 2001, ApJ, 548, L175
\bibitem[\protect\citeauthoryear{Cowley \& Schmidtke}{1990}]{cowley} Cowley, A. P., Schmidtke, P. C. 1990, ApJ, 99, 678
\bibitem[\protect\citeauthoryear{Deutsch et al.}{1996}]{deut} Deutsch, E. W., Margon, B., Wachter, S., Anderson, S. F. 1996, ApJ, 471, 979 
\bibitem[\protect\citeauthoryear{Forman et al.}{1978}]{form} Forman, W., Jones, C., Cominsky, L., et al. 1978, ApJ, 38, 357
 \bibitem[\protect\citeauthoryear{Garcia \& Grindlay}{1987}]{gar87} Garcia, M. R., Grindlay, J. E. 1987, ApJ, 313, L59
\bibitem[\protect\citeauthoryear{Garcia et al.}{1989}]{gar89} Garcia, M. R., Bailyn, C. D., Grindlay, J. E., Molnar, L. A. 1989, ApJ, 341, L75
\bibitem[\protect\citeauthoryear{Garcia et al.}{1992}]{gar92} Garcia, M.; Grindlay, J., Bailyn, C. 1992, IAUC, 5578, 2 
\bibitem[\protect\citeauthoryear{Garcia}{1994}]{gar94} Garcia, M. R. 1994, ApJ, 435, 407
\bibitem[\protect\citeauthoryear{Garcia \& Callanan}{1999}]{gar99} Garcia, M. R., Callanan, P. J. 1999, ApJ, 118, 1390
\bibitem[\protect\citeauthoryear{Heinz \& Nowak}{2001}]{heinz} Heinz, S., Nowak, M. A. 2001, MNRAS, 320, 249
\bibitem[\protect\citeauthoryear{Horne et al.}{1986}]{horne} Horne, K., Verbunt, F., Schneider, D. P. 1986, MNRAS, 218, 63
\bibitem[\protect\citeauthoryear{Jansen et al.}{2001}]{j01} Jansen, A., Lumb, D., Altieri B., et al. 2001, A\&A, 365, L1
\bibitem[\protect\citeauthoryear{Kaluzny}{1988}]{kal88} Kaluzny, J. 1988, Acta Astronomica, 38, 207
\bibitem[\protect\citeauthoryear{Mangano et al.}{2004}]{Mang} Mangano, V., Israel, G. L., Stella, L. 2004, A\&A, 419, 1045
\bibitem[\protect\citeauthoryear{McClintock et al.}{1981}]{mc81} McClintock, J. E., Remillard, R. A., Margon, B. 1981, ApJ, 243, 900 (MC81)
\bibitem[\protect\citeauthoryear{McClintock et al.}{1982}]{mc82} McClintock, J. E., London, R. A., Bond, H. E., Grauer, A. D. 1982, ApJ, 258, 245 (MC82)
\bibitem[\protect\citeauthoryear{Nowak et al.}{2002}]{now} Nowak, M. A., Heinz, S., Begelman, M. C. 2002, ApJ, 573, 778 (N02)
\bibitem[\protect\citeauthoryear{Papitto et al.}{2005}]{pap} Papitto, A., Menna, M. T., Burderi, L., et al 2005, ApJ, 621, L113
\bibitem[\protect\citeauthoryear{Parmar et al.}{2000}]{parm} Parmar, A. N., Oosterbroek, T., Del Sordo, S., et al. 2000, A\&A, 356, 175
\bibitem[\protect\citeauthoryear{Parmar et al.}{1991}]{parmar91} Parmar, A. N., Smale, A. P., Verbunt, F., Corbet, R. H. D. 1991, ApJ, 366, 253
\bibitem[\protect\citeauthoryear{Pietsch et al.}{1986}]{pietsch86} Pietsch, W., Steinle, H., Gottwald, M., Graser, U. 1986, A\&A, 157, 23
\bibitem[\protect\citeauthoryear{Stella et al.}{1994}]{stella94} Stella, L., Campana, S., Colpi, M., Mereghetti, S., Tavani, M.
 1994, ApJ, 423, L47
\bibitem[\protect\citeauthoryear{Thorstensen et al.}{1979}]{Th79} Thorstensen, J., Charles, P., Bowyer, S., et al. 1979, ApJ, 233, L57 (Th79)
\bibitem[\protect\citeauthoryear{Thorstensen et al.}{1988}]{Th88} Thorstensen, J. R., Brownsberger, K. R., Mook, 
D. E., et al. 1988, ApJ, 334, 430
\bibitem[\protect\citeauthoryear{Ulmer et al.}{1980}]{ulm} Ulmer, M. P., Shulman, S., Yentis, D., et al. 1980, ApJ, 235, L159
\bibitem[\protect\citeauthoryear{Verbunt}{1993}]{verbunt} Verbunt, F. 1993, ARA\&A, 31, 93 
\bibitem[\protect\citeauthoryear{Wenzel et al.}{1983}]{wen83} Wenzel, W. 1983, IBVS, 2452, 1
\bibitem[\protect\citeauthoryear{White \& Holt} {1982}]{wh82} White, N. E., Holt, S. S. 1982, ApJ, 257, 318
\bibitem[\protect\citeauthoryear{White et al.}{1995}]{pnw} White, N. E., Nagase, F., Parmar, A. N., in X-ray Binaries, ed. Lewin,
W. H. G., Van Paradijs J., van den Heuvel, E. P. J., Cambridge University Press 1995, p. 1
\bibitem[\protect\citeauthoryear{Wijnands}{2002}]{win} Wijnands, R. 2002, ASPC, 262, 235 
\bibitem[\protect\citeauthoryear{Zavlin et al.}{1996}]{zavlin} Zavlin, V. E., Pavlov, G. G., Shibanov, Y. A. 1996, A\&A, 315,141
\end{thebibliography}
\end{document}